\documentclass[11pt]{article}
\usepackage{color}
\usepackage{graphicx}

\newcommand{\be}{\begin{displaymath}}
\newcommand{\bn}{\begin{equation}}

\newcommand{\en}{\end{equation}}
\newcommand{\ee}{\end{displaymath}}
\newcommand{\p}{\partial}
\newcommand{\lang}{\left\langle}
\newcommand{\rang}{\right\rangle}

\begin{document}
\begin{center}

\Large

{\bf Constraints on dynamo action in plasmas}
~\\
\normalsize

\vspace{1cm}
P. Helander$^{1,2,3}$, M. Strumik$^2$ and A.A. Schekochihin$^{2,3}$
\\[1cm]
{\it $^1$Max-Planck-Institut f\"ur Plasmaphysik, 17491 Greifswald, Germany \\
$^2$ Rudolf Peierls Centre for Theoretical Physics, University of Oxford, \\ Oxford OX1 3NP, United Kingdom \\
$^3$ Merton College, Oxford OX1 4JD, United Kingdom}

\vspace{0.5cm}

\end{center}

\noindent

Upper bounds are derived on the amount of magnetic energy that can be generated by dynamo action in collisional and collisionless plasmas with and without external forcing. A hierarchy of mathematical descriptions is considered for the plasma dynamics: ideal MHD, visco-resistive MHD, the double-adiabatic theory of Chew, Goldberger and Low (CGL), kinetic MHD, and other kinetic models. It is found that dynamo action is greatly constrained in models where the magnetic moment of any particle species is conserved. In the absence of external forcing, the magnetic energy then remains small at all times if it is small in the initial state. In other words, a small ``seed'' magnetic field cannot be amplified significantly, regardless of the nature of flow, as long as the collision frequency and gyroradius are small enough to be negligible. A similar conclusion also holds if the system is subject to external forcing as long as this forcing conserves the magnetic moment of at least one plasma species and does not greatly increase the total energy of the plasma (i.e., in practice, is subsonic). Dynamo action therefore always requires collisions or some small-scale kinetic mechanism for breaking the adiabatic invariance of the magnetic moment.

\newpage

\section{Introduction}

Dynamo action is believed to be ubiquitous throughout the Universe. It is generally agreed that in planetary and stellar interiors as well as in the interstellar and intergalactic medium, turbulent fluid motions are responsible for the presence of magnetic fields \cite{Zweibel,Brandenburg,Kulsrud-Zweibel,Durrer,Roberts}, i.e., pre-existing ``seed'' fields have been amplified to their present level and are now maintained by dynamo action. The precise mechanism by which this happens is, however, still a matter of debate. It is therefore of interest to identify the conditions under which dynamo action is (im)possible. 

The great majority of all research into dynamo theory has been carried out within the framework of magnetohydrodynamics (MHD), although most interplanetary, interstellar and intergalactic plasmas are too weakly collisional to be accurately described by this approximation. In the present paper, we investigate how dynamo action is different in mathematical plasma models that go beyond MHD and capture some of the underlying kinetic dynamics of collisionless plasmas. In order to isolate this consideration from the question of how the geometry of the flow or external forces affect the dynamo, we first consider the free evolution (i.e., without external forces) of a conducting fluid from an arbitrary state at time $t=0$ and ask to what extent the resulting motion can amplify a pre-existing magnetic field. This question is meaningful even for an ideally conducting fluid, where steady-state dynamo action would inevitably give rise to magnetic fields at infinitely small scales (see, e.g., \cite{Zeldovich}). In general, an arbitrarily prepared state at $t=0$ is not in equilibrium, and the ensuing motion will be chaotic, stretching and bending magnetic field lines, thereby increasing the magnetic field energy, even in the absence of resistivity and reconnection.\footnote{In a steady-state situation, the question of taking these effects into account is usually about whether the field can survive in their presence \cite{Zeldovich,Chertkov,Schekochihin-2004}. The basic {\em amplification} mechanism for turbulent, small-scale dynamo is ideal and relies on chaotic stretching of field lines.} But how much can the magnetic energy increase above its initial value? To what extent can the thermal and kinetic energy of the fluid be converted into magnetic energy? 

This problem was first considered in an early paper by Batchelor \cite{Batchelor}, who used the visco-resistive MHD equations and considered homogeneous and isotropic turbulence as the underlying fluid flow. He concluded that, if the conductivity is large and the initial magnetic field weak, the magnetic energy will in general grow exponentially with time until ``the large wave-number components contain comparable amounts of kinetic and magnetic energy''. In a contemporaneous article, Biermann and Schl\"uter  \cite{Biermann} similarly concluded that ``the magnetic energy density will finally reach the energy density of the turbulence''.\footnote{Although their view was that it would be the {\em total} energy density of the turbulence that the magnetic energy would become comprabale to --- this indeed appears to be the case \cite{Haugen,Beresnyak}.} While there was then, and is now, some level of disagreement or uncertainty about the detailed state resulting from turbulent dynamo action in MHD, simplified analytical models \cite{Zeldovich,Kazantsev,Chertkov,Boldyrev} and numerical simulations \cite{Meneguzzi,Schekochihin-2004,Haugen,Beresnyak} as well as recent laboratory experiments \cite{Monchaux,Tzeferacos} demonstrate beyond reasonable doubt that a sufficiently chaotic three-dimensional flow of a conducting MHD fluid at large enough magnetic Reynolds numbers will generate tangled magnetic fields with energy densities comparable to that of the fluid motions. 

The situation can be very different in mathematical models other than MHD, as was first realised by Kulsrud et al. \cite{Kulsrud-1997} who pointed out that conservation of the first adiabatic invariant (magnetic moment) for each gyrating particle implied that changing the magnetic field strength by a finite factor would require changing the perpendicular energy of the particles (i.e., the perpendicular pressure) by a similar factor, which is usually not possible due to energy-source constraints.\footnote{They further argued that in the presence of pressure anisotropy, which would that arise from any local increase of the magnetic-field strength, the magnetic mirror force might modify the velocity field in such a way as to prevent any further field amplification, independently of the absolute magnitude of the field (i.e., even for dynamically weak fields). The microphysical feedback on the macroscopic motions remains poorly understood and a subject of much current interest \cite{Kunz,Melville,Riquelme-2016,Squire}.} In this work, we extend the arguments of Kulsrud et al. \cite{Kulsrud-1997} by exploring a hierarchy of plasma models of increasing complexity, starting from ideal and resistive MHD and proceeding to the double-adiabatic theory of Chew, Goldberger and Low (CGL) \cite{CGL}, kinetic MHD \cite{Kulsrud}, and more complete kinetic models. In each case, it is possible to derive a rigorous upper bound on the magnetic energy that is valid at all times, and this bound turns out to depend crucially on the conservation properties of the equations, particularly on the equation of state. In all models appropriate for a collisionless plasma with vanishingly small gyroradii, the growth of the magnetic energy is severely limited. Having established these results, we introduce external forcing and show that the constraints found on the growth of magnetic energy are still valid if the energy input from the external forces is not too large. These conclusions are tested in numerical simulations of the CGL equations (cf. \cite{Lima}). The conclusion is that no dynamo action is possible within any model of a gyrotropic plasma that does not allow the conservation of the first adiabatic invariant of any species to be broken. 

\section{MHD and CGL equations}

We first consider the question whether the flow of a conducting fluid can cause amplification of a ``seed'' magnetic field when evolving freely from an initial state according to the inviscid, ideal MHD equations \cite{Kulsrud}
	\bn \frac{d \rho}{dt} + \rho \nabla \cdot {\bf V} = 0, \label{conteq} \en
	\bn \rho \frac{d {\bf V}}{dt} = {\bf j} \times {\bf B} - \nabla \cdot {\sf P}, \label{momeq} \en
	\bn \frac{\partial {\bf B}}{\partial t} = \nabla \times ( {\bf V} \times {\bf B}), \label{indeq} \en
	\bn \nabla \times {\bf B} = {\bf j}, \label{Ampere} \en
where the symbols have their usual meaning, $d/dt = \p / \p t + {\bf V} \cdot \nabla$ denotes the convective derivative, and the pressure tensor is gyrotropic,
	\bn {\sf P} = p_\perp {\bf I} + (p_\| - p_\perp) {\bf bb}, \label{P} \en
with ${\bf b} = {\bf B}/B$. The components of $\sf P$ are determined either by ideal MHD entropy conservation,
	$$ p_\perp = p_\| = p, $$
	\bn \frac{d}{dt} \left( \frac{p}{\rho^\gamma} \right) = 0, 
	\label{ideal MHD}
	\en
where $\gamma = 5/3$, or by the CGL equations \cite{CGL}
	\bn \frac{d}{dt} \left( \frac{p_\perp}{\rho B} \right) = 0, \label{1} \en
	\bn \frac{d}{dt} \left( \frac{p_\| B^2}{\rho^3} \right) = 0. \label{2} \en
The adiabatic law (\ref{ideal MHD}) holds in this latter case, too, if it is understood that $p$ is formally replaced by $(p_\perp^2 p_\|)^{1/3}$. 	
	
The boundary conditions are either periodic in all three directions or it is assumed that the fluid is surrounded by a rigid, ideally conducting boundary, at which the normal components of $\bf V$ and $ {\bf B}$ vanish. 
In both cases, and for both the MHD and the CGL models, the total energy, defined by
	\bn W = \lang \frac{\rho V^2}{2} + p_\perp + \frac{p_\|}{2} + \frac{B^2}{2} \rang, 
	\label{W}
	\en
 is then conserved, where the angular brackets denote a volume average.

\subsection{Available energy in ideal MHD} \label{sec_A_MHD}

Eliminating the density from the continuity equation (\ref{conteq}) and the adiabatic law (\ref{ideal MHD}) gives
	$$ \frac{\p p^{1/\gamma}}{\p t} + \nabla \cdot \left( p^{1/\gamma} {\bf V} \right) = 0, $$
which implies that the quantity
	$$ S = \lang p^{1/\gamma} \rang $$
is conserved, $d S / dt = 0$. We thus have two conserved quantities, and may seek the maximum of the magnetic energy
	$$ M = \lang \frac{B^2}{2} \rang $$
under the constraint of constant $W$ and $S$. Since $M < W$, it is clear that this maximum indeed exists and represents a mathematical upper bound on the magnetic energy at all times $t\ge 0$. It may, however, be inaccessible from the initial conditions, because there is an infinity of other constraints associated with the topology of the magnetic field lines \cite{Taylor-Newton}, but we ignore this issue as we only seek an upper bound on $M$ that need not be the best possible one. 

Thus, we consider the maximum of the functional
	$$ T[\rho,{\bf V},{\bf B}, p; \lambda, \mu] = M - \lambda (W-W_0) - \mu (S-S_0), $$
where $\lambda$ and $\mu$ are Lagrange multipliers, and initial conditions are denoted by a $0$ subscript. Since this functional does not involve any derivatives of the fields $(\rho, {\bf V}, {\bf B}, p)$, these must be constant at the maximum in question; this becomes obvious if one writes down the Euler-Lagrange equations for the minmising fields. Denoting these quantities by the subscript $1$, we thus have
	$$ S = \lang p_0^{1/\gamma} \rang = p_1^{1/\gamma} $$
and $V_1 = 0$, so that
	$$ M_1 = W - \frac{\lang p_0^{1/\gamma} \rang^\gamma}{\gamma-1}. $$
The difference between the upper bound $M_1$ and the intial value of the magnetic energy, $M_0$, is thus
	\bn A = M_1 - M_0 = \frac{\lang p_0 \rang- \lang p_0^{1/\gamma} \rang^\gamma}{\gamma-1} + \lang \frac{\rho_0 V_0^2}{2} \rang, 
	\label{A}
	\en
This quantity is always positive because, for $\gamma > 1$,
	$$ \lang p_0^{1/\gamma} \rang \le \lang p_0 \rang^{1/\gamma} $$
by H\"older's inequality. 

The expression (\ref{A}) represents an upper bound on the amount of thermal and kinetic energy that is available for conversion into magnetic energy, and will in the following be referred to as the ``available energy'' for dynamo action. This upper bound is usually inaccessible: not only does it ignore topological constraints on the evolution of the magnetic field, but it is also in general incompatible with the boundary conditions. As we shall see, it can nevertheless be surprisingly restrictive.  

If the initial state is such that either the Mach number is of order unity or the pressure $p_0$ varies substantially (by order unity) across the domain, then a substantial fraction of the thermal and kinetic energy is available for conversion into magnetic energy, $A/W = O(1)$. If, however, the initial pressure fluctuations are small,
	$$ p_0({\bf r}) = P + \delta p({\bf r}), $$                                        
where $P = \lang p_0 \rang$ and $\delta p \ll P$, then the available energy from these fluctuations is quadratic in their amplitude and is relatively small:
	\bn A_{\delta p} = \frac{\lang \delta p^2 \rang}{2 \gamma P} \ll W.
        \label{A_dp}
        \en
If we consider specifically the limit of small Mach numbers (subsonic motions): 
	\bn {\rm Ma} \equiv \frac{V_0}{\sqrt{\gamma p_0/\rho_0}} \ll 1, 
        \label{small_Ma}
        \en
the dynamics are pressure-balanced and, typically, pressure perturbations are $\delta p/p_0 \sim {\rm Ma}^2$. This means that, to lowest order in ${\rm Ma}$, the available energy is just the {\em kinetic} energy in the initial state, $K_0 = \lang \rho_0 V_0^2/2 \rang$: 
	\bn \frac{A}{K_0} = 1 + {\cal O}({\rm Ma}). 
        \label{A_subsonic}
        \en                                
	
	\subsection{Finite resistivity and viscosity}\label{sec_diss}

Most dynamos considered in the literature involve resistivity, which enables field-line diffusion and reconnection to take place. A justified question is, therefore, to what extent the above conclusions remain valid if a finite resistivity is introduced. The induction equation (\ref{indeq}) then becomes 
	$$ \frac{\partial {\bf B}}{\partial t} = \nabla \times ( {\bf V} \times {\bf B}) + \eta \nabla^2 {\bf B}, $$ 
and the entropy conservation law (\ref{ideal MHD}) is replaced by 
	\bn \frac{d}{dt} \left( \frac{p}{\rho^\gamma} \right) = \frac{(\gamma - 1) \eta j^2}{\rho^\gamma}. 
	\label{entropy production}
	\en
The energy $W$ is still conserved, but the quantity $S = \lang p^{1/\gamma} \rang$ now increases with time
	$$ S(t) > S_0, $$
reflecting positive entropy production. However, regardless of how much $S$ increases, we may still seek the maximum magnetic energy $M_1$ at a given $W$ and $S=S_1$ and thus find an upper bound on the available energy 
	\bn A = M_1 - M_0 = \frac{\lang p_0 \rang- S_1^\gamma}{\gamma-1} + \lang \frac{\rho_0 V_0^2}{2} \rang. 
        \label{A_diss}
        \en
Since $S_1 > S_0$, this bound is lower than in the ideal-MHD case, and we conclude that less energy is available for conversion to magnetic energy. In other words, the bound (\ref{A}) still holds, but is even tighter than before. 

The addition of viscosity to the MHD equations has a similar effect, with viscosity now contributing to entropy production (\ref{entropy production}). Since the total energy (\ref{W}) is conserved and $S(t)$ increases with time, the magnetic energy is again bounded from above by Eq.~(\ref{A}). 
	
\subsection{Available energy in the double-adiabatic model} \label{sec_A_CGL}

As we shall now see, the situation is very different in the double-adiabatic model of Chew, Goldberger and Low \cite{CGL,Kulsrud}. Eliminating the density from the continuity equation (\ref{conteq}) and from the equations of state (\ref{1}) and (\ref{2}) gives
	$$ \frac{\p }{\p t} \left( \frac{p_\perp}{B} \right) + \nabla \cdot \left( \frac{p_\perp {\bf V}}{B} \right) = 0, $$
	$$ \frac{\p }{\p t} \left( p_\|^{1/3} B^{2/3} \right) + \nabla \cdot \left( p_\|^{1/3} B^{2/3} {\bf V} \right) = 0, $$
implying that the following two quantities are conserved\footnote{Additional invariants are derived in Appendix A, but we do not consider the additional constraints implied by their existence.}
	\bn I = \lang \frac{p_\perp}{B} \rang, 
	\label{I}
	\en
	$$ J = \lang p_\|^{1/3} B^{2/3} \rang. $$
We may thus ask for the maximum of the magnetic energy $M$ under the constraint that $W$, $I$ and $J$ are fixed. Proceeding as before, we find that all the fields $(p_\perp, p_\|, B, \ldots)$ are constant and the flow velocity vanishes, so in the state of maximum magnetic energy we have
	$$ W = IB_1 + \frac{J^3}{2 B_1^2} + \frac{B_1^2}{2}. $$
This equation is a quartic for $B_1$ as a function of the invariants $I$, $J$, and $W$, and in terms of the normalised magnetic field $b = B_1 / \sqrt{W}$, it becomes
	\bn b^4 + \frac{2Ib^3}{\sqrt{W}}  - 2 b^2 + \frac{J^3}{W^2} = 0. 
	\label{quartic}
	\en
The sum of all four roots is equal to $-2 I / \sqrt{W} < 0$, while their product equals $J^3/W^2 > 0$. There are, therefore, two positive and two negative roots. The negative ones can obviously be discarded, and the largest positive root is the upper bound that we are seeking. The other positive root is a {\em lower} bound on the magnetic energy, for the constancy of $J$ implies that $B$ cannot be made arbitrarily small at constant $W$, since small $B$ implies large $p_\|$. This is different from the ideal-MHD case, where the analogous calculation yields no lower bound on $M$ other than $B=0$. 

In the intial state, the ratio of thermal to magnetic energy is equal to 
		$$ \frac{3\beta_0}{2} = \frac{\lang p_{\perp 0} + p_{\|0}/2 \rang}{M_0} = \frac{W - \rho_0 V_0^2/2}{M_0} - 1, $$
and in the state of maximum magnetic energy it is 
	$$ \frac{3 \beta_1}{2} = \frac{W}{M_1} - 1 = \frac{2}{b^2} - 1. $$
The general solution of Eq.~(\ref{quartic}) is not particularly edifying, but it does yield interesting information in the limit of high $\beta_0$. In this limit,
		$$ \frac{I}{\sqrt{W}} \sim \sqrt{\beta_0} \gg 1, $$
		$$ \frac{J^3}{W^2} \sim \frac{1}{\beta_0} \ll 1, $$
and it is helpful to write $x = b \sqrt{\beta_0}$, so that Eq.~(\ref{quartic}) becomes
	\bn \frac{x^4}{\beta_0} + 2 a x^3 - 2 x^2 + c = 0, 
        \label{x_eqn}
        \en
where $a = I / \sqrt{\beta_0 W}$ and $c = \beta_0 J^3 / W^2$ are of order unity. Since $\beta_0$ is large, one of the negative roots is obtained by balancing the first two terms, $x = - 2 a \beta_0$, whereas the three remaining roots are of order unity and are found by neglecting the first term. We thus come to the conclusion that if the initial state is one with little magnetic field, so that $\beta_0$ is large, then $\beta$ will remain large at all times. Indeed, $\beta$ can only change by order unity {\em in either direction} because there is a lower as well as an upper bound on the magnetic energy, both of order $b^2 \sim 1/\beta_0$ since $x = O(1)$. In other words, in the high-$\beta$ limit, the available energy is a small fraction of the total energy,
	\bn A \sim \frac{W}{\beta_0}, \label{CGL limit}
	\en
in contrast to conventional MHD, where $A \sim W$. 

In the small-Mach-number-limit (\ref{small_Ma}), the fact that only a small fraction of the total energy is available to be converted into magnetic field is not by itself very surprising and it is relevant to ask how much of the {\em kinetic} energy in the initial state can be converted into magnetic energy. Since
	\bn \frac{A}{K_0} \sim \frac{1}{{\rm Ma}^2 \beta_0}, 
        \label{A_CGL_subsonic}
        \en
only a small fraction of $K_0$ is available for conversion if $\beta_0 \gg {\rm Ma}^{-2}$. This is in contrast to MHD, where all of $K_0$ is available: see (\ref{A_subsonic}).

\section{Available energy in kinetic plasma models}

The upper bound on the magnetic energy that we have derived arises because the invariant $I$ dictates that the magnetic field cannot be increased without a similar {\em relative} increase in $p_\perp$, just as anticipated in \cite{Kulsrud-1997}. Thus, even if the magnetic field is very weak, it ``costs'' a significant amount of energy to increase it by a finite factor. This property of the CGL equations only relies on the constancy of the magnetic moment and is, therefore, shared by any plasma model that conserves this quantity.

\subsection{Kinetic MHD} \label{sec_A_KMHD}

An example of such a model is kinetic MHD, which is obtained by expanding the Vlasov-Maxwell system of equations in the small-gyroradius limit, ordering the plasma flow velocity to be at most comparable to the ion thermal speed, $V \sim v_{Ti}$ \cite{Kulsrud,Hazeltine}. The fluid equations then obtained coincide with our Eqs.~(\ref{conteq}) - (\ref{P}), but the components $(p_\perp, p_\|)$ of the pressure tensor are determined by the kinetic equation
	\bn \frac{\p f_s}{\p t} + \left( v_\| {\bf b} + {\bf V} \right) \cdot \nabla f_s + \dot \epsilon \frac{\p f_s}{\p \epsilon} = 0, 
	\label{dke}
	\en
	\bn \dot \epsilon = e_s v_\| E_\| - m_s v_\| {\bf b} \cdot \frac{´d {´\bf V}}{dt} - \mu B \nabla \cdot {\bf V} - (m_s v_\|^2 - \mu B) {\bf bb} : \nabla {\bf V}, 
	\label{epsilon}
	\en
rather than by Eqs.~(\ref{1}) and (\ref{2}) (which follow from kinetic MHD only if heat fluxes are neglected). Here, $f_s$ denotes the distribution function of the particles of species $s$, whose charge is denoted $e_s$ and mass by $m_s$, the magnetic moment is $\mu = m_s v_\perp^2/2B$, and the particle velocity $\bf v$ is measured relative to the mean velocity ${\bf V}({\bf r}, t)$, so that the laboratory-frame velocity of a particle is ${\bf u} = {\bf V} + {\bf v}$. From the solution of the kinetic equation (\ref{dke}), the pressures needed in the equation of motion (\ref{momeq}) are computed by
	$$ {p_\perp \choose p_\|} = \sum_s \int {\mu B \choose m_s v_\|^2 } f_s \; d^3v. $$
Another difference with conventional MHD is that $E_\|$, the component of the electric field that is parallel to $\bf B$, appears in Eq.~(\ref{epsilon}). As in MHD, $E_\|$ is relatively small, as follows from the observation that the first term on the right-hand side of Eq.~(\ref{epsilon}) is comparable to the others when $E_\| / B \sim \rho_i v_{Ti} / L$, where $\rho_i = m_i v_{Ti}/e_iB$ is the ion gyroradius. However, unlike in MHD, $E_\|$ affects the motion of the plasma and must be determined by the quasineutrality condition 
	$$ \sum_s e_s \int f_s \; d^3v = 0, $$
which closes the kinetic-MHD system of equations. 

This system conserves both the energy (\ref{W}) and the total magnetic moment (\ref{I}), the latter for each species individually. This is most easily shown by first writing the kinetic equation (\ref{dke}) in conservative form,
	\bn \frac{\p}{\p t} \left( \frac{B f_s}{v_\|} \right) + \nabla \cdot \left[ \left( {\bf B} + \frac{B {\bf V}}{v_\|} \right) f_s \right] 
	+ \frac{\p}{\p \epsilon} \left( \frac{\dot \epsilon B f_s}{v_\|} \right) = 0. 
	\label{conservative form}
	\en
Multiplying this equation by
	$$ \frac{\mu v_\|}{B} \; d^3v = \mu \sum_{\sigma} \frac{2 \pi}{m_s^2} d\mu d\epsilon, $$
where $\sigma = v_\| / |v_\| |$, and integrating over velocity and real space gives $dI_s/dt = 0$. Obviously, the total magnetic moment
	$$ I = \sum_s I_s $$
is also conserved. To prove energy conservation, one multiplies Eq.~(\ref{conservative form}) by $(\epsilon v_\| /B) \; d^3v$ integrates similarly, and sums over species, giving
	$$ \frac{d}{dt} \lang p_\perp + \frac{p_\|}{2} \rang = \lang j_\| E_\| - p_\perp \nabla \cdot {\bf V} + (p_\perp - p_\|) {\bf bb} : \nabla {\bf V} \rang, $$
where
	$$ j_\| = \sum_s e_s \int v_\| f_s d^3v = 0 $$
to the requisite order  \cite{Kulsrud}. The evolution of the kinetic energy can be computed from the continuity and momentum equations (\ref{conteq})-(\ref{momeq}), 
	$$ \frac{d}{dt} \lang \frac{\rho V^2}{2} \rang 
	= \lang {\bf V} \cdot ({\bf j} \times {\bf B}) + p_\perp \nabla \cdot {\bf V} - (p_\perp - p_\|) {\bf bb} : \nabla {\bf V} \rang, $$
and the magnetic energy evolves according to
	$$ \frac{d}{dt} \lang \frac{B^2}{2} \rang = \lang ({\bf V} \times {\bf B} ) \cdot {\bf j} \rang, $$
as found from the induction equation (\ref{indeq}). The sum of these energy relations implies that the total energy (\ref{W}) is conserved. 

Knowing that $W$ and $I$ are conserved, we again proceed to seek the state of maximum magnetic energy 
	$$ M = W - L - \lang p_\perp \rang, $$
where we have denoted 
	$$ L = \lang \frac{\rho V^2}{2} + \frac{p_\|}{2} \rang. $$
As before, this state has constant magnetic field strength, so $M = B_1^2/2$ and
	$$ B_1^2 + 2 I B_1 - 2(W-L) = 0, $$
i.e.,
	$$ B_1 = I \left( \sqrt{ 1 + \frac{2 (W-L)}{I^2}} - 1 \right). $$
At high beta, $W \ll I^2$, we conclude that $B_1 \simeq (W-L)/I$, and the maximum magnetic energy is a small fraction of the total energy, 
	\bn \frac{M_1}{W} = \frac{(W-L)^2}{2 W I^2} \le \frac{W}{2I^2} \sim \frac{1}{\beta_0} \ll 1. 
	\label{bound in kinetic MHD}
	\en
As in the case of the CGL model (to which the present argument also applies), if the Mach number of the initial flow is small, the fraction of kinetic energy that can be converted to magnetic energy is
	$$ \frac{M_1}{K_0} \sim \frac{1}{{\rm Ma}^2 \beta_0}, $$
and is small at high enough $\beta_0$. This is the same result as (\ref{A_CGL_subsonic}). 
	
\subsection{More general plasma models} \label{sec_A_other}

The analysis leading to the bound (\ref{bound in kinetic MHD}) on the magnetic energy shows that it is the conservation of the magnetic moment that leads to the available energy being so limited at high beta. The limit (\ref{bound in kinetic MHD}) is thus applicable beyond the approximations made in kinetic MHD and will hold in any plasma model where the total magnetic moment and energy are conserved and the latter can be written in the form
	\bn W = L + \lang p_\perp \rang + M, 
	\label{general W}
	\en
where $L$ is a positive definite quantity. Mathematically, the argument is exactly the same as that just given for kinetic MHD. Any kinetic description of the plasma in which the gyroradius is small, all frequencies are lower than the ion cyclotron frequency, and collisions are negligible has this property. Drift kinetics, which is similar to kinetic MHD but treats the plasma flow velocity as smaller than the ion thermal speed is an example \cite{Hazeltine},\footnote{This is the kind of ordering that can, for example, be argued to be appropriate for the turbulent plasma in galaxy clusters \cite{Rosin}.} under the proviso that collisions can be ignored, so that the total magnetic moment $I$ is conserved. The magnetic field can therefore only grow appreciably on time scales longer than the collision time, at which point $I$ is no longer conserved. 

The perpendicular pressure $p_\perp$ entering in Eq.~(\ref{general W}) and the definition (\ref{I}) of $I$ need not refer to the entire plasma, but could denote just the perpendicular pressure of one of its components. For the bound (\ref{bound in kinetic MHD}) to hold, it is sufficient that the magnetic moments for the particles of {\em one} plasma component be conserved (as long as its number density is not so small as to render the high-$\beta$ approximation invalid). For instance, if we consider dynamo action on time scales that are longer than the collision time for electrons but shorter than that for ions, the collision operator needs to be retained in the electron kinetic equation but can be ignored in the ion dynamics. If, in this situation, the initial ion beta is large and $p_\perp$ denotes the perpendicular ion pressure, the bound (\ref{bound in kinetic MHD}) implies that the magnetic energy cannot grow significantly, even though the conservation of the magnetic moment of the electrons is broken by collisions. 

\section{Case of external forcing}

\subsection{MHD model} \label{sec_forced_MHD}

Numerical dynamo simulations usually involve external forcing \cite{Meneguzzi,Schekochihin-2004,Haugen,Beresnyak,Lima}, and we now ask how this may affect our results. If a force is added to the equation of motion, 
	$$ \rho \frac{d {\bf V}}{dt} = {\bf j} \times {\bf B} - \nabla \cdot {\sf P} + {\bf F}({\bf r}, t), $$
the total energy $W$ is no longer conserved, but increases (or decreases), as the force adds kinetic energy to the system. In a dissipative plasma, this energy is continually processed: it is transferred to thermal energy by viscosity and resistivity.\footnote{It can also then be radiated, possibly leading to a statistically constant $W$, as, for example, is believed to happen in galaxy-cluster cores \cite{Zhuravleva}. Examining further arguments in this section, one might expect this approximate conservation of $W$ to help establish better upper bounds on the attainable magnetic energy. However, in subsonically turbulent systems, this steady state is achieved on the heating/cooling time scales, which, as we will argue in what follows, are much longer than the dynamical time scales on which dynamo action matters.} Since entropy is constantly produced by these dissipation processes, $S$ is also no longer conserved. If, over some period of time, the total energy increases from $W_0$ to $W_1$ and entropy from $S_0$ to $S_1$ and we maximise $M_1$ at given values of $W_1$ and $S_1$, we find (similarly to the calculation in section \ref{sec_diss})
	$$ M_1 = W_1 - \frac{S_1^\gamma}{\gamma - 1}, $$
and the available energy becomes
	\bn A = M_1 - M_0 = W_1 - W_0 
	+ \frac{\lang p_0 \rang- S_1^\gamma}{\gamma-1} + \lang \frac{\rho_0 V_0^2}{2} \rang. 
        \label{A_forced}
        \en
Compared to the ideal, unforced case (\ref{A}), this makes additional energy available for conversion into magnetic energy (the work done by the external force can go into magnetic energy), attenuated, as in (\ref{A_diss}),  by the fact that $S_1 > S_0 = \lang p_0^{1/\gamma}\rang$ (continuous forcing will produce motions and magnetic fields that will eventually reach dissipative scales and be thermalised, producing entropy). If $F$ is arbitrary, the above calculation does not establish any bound on the increase of the magnetic energy. 

Considering again the low-Mach-number limit (\ref{small_Ma}), we can, as we did at the end of section \ref{sec_A_MHD}, neglect the term in (\ref{A_forced}) that contains $S_1$ and can be bounded from above by (\ref{A_dp}) because $S_1>S_0$. For simplicity, let us assume an initially motionless state ($V_0=0$), which, for a forced system, does not restrict generality in any significant way. If external forcing injects energy into the system at the mean rate $\varepsilon$, then the available energy after time $t$ is 
	$$ A \approx W_1 - W_0 = \varepsilon t. $$
In a subsonically forced and, therefore, low-Mach-number system, the time that it takes the external forcing to inject an amount of energy comparable to the total energy is asymptotically long. 
If the {\em kinetic} energy of the plasma flows at time $t$ is $K_1 \sim {\rm Ma}^2 W_1$, we may define the ``dynamical'' time as the typical time over which external forcing can build up motions with this energy, $\tau_{\rm dyn} = K_1/\varepsilon$. The total energy can only change by an amount of order unity after the ``heating'' time $t\sim W_1/\varepsilon\sim \tau_{\rm dyn}/{\rm Ma}^2 \gg \tau_{\rm dyn}$. The interesting question in the context of dynamo action in such a forced system is how much energy is available for conversion into magnetic fields over times of order $\tau_{\rm dyn}$, which immediately gives us 
	$$ A \sim \varepsilon \tau_{\rm dyn} \sim K_1. $$
This is just a (perhaps overcomplicated) way of stating that in subsonic MHD turbulence, over dynamical times, a possible dynamo mechanism has at its disposal energies of order the kinetic energy of the motions, a result analogous to (\ref{A_subsonic}). 

\subsection{CGL and other kinetic plasma models} \label{sec_forced_CGL}

Let us assume that the external forcing does not directly break adiabatic invariance, i.e., that it occurs on sufficiently long scales in time and spaces. Nevertheless, the dissipation of the injected energy will in general occur on small scales and via processes (e.g., collisional viscosity and resistivity) that do not conserve $I$ and $J$. Thus, just like in the case of the MHD model, we are left without exact conservation laws that would allow us to constrain dynamo action. Let us, however, again focus on the limit of low Mach numbers and consider the evolution of the system over dynamical (rather than heating) time scales. Over such times, both the total energy of the plasma and the CGL invariants can only change by small amounts: at most,  
        $$ W_1 - W_0 \sim \varepsilon \tau_{\rm dyn} \sim {\rm Ma^2} W_1,$$
        $$ I_1 - I_0 \sim \frac{\varepsilon \tau_{\rm dyn}}{B_0} \sim {\rm Ma}^2\sqrt{\beta_0 W_0},$$
and similarly for $J$. This implies that if we seek to maximise magnetic energy in a subsonically forced CGL model subject to some fixed values of $W_1$, $I_1$ and $J_1$, the argument presented in section \ref{sec_A_CGL} continues to be valid, with the coefficients $a$ and $c$ in (\ref{x_eqn}) still of order unity because they only differ by ${\cal O}({\rm Ma}^2)$ from the values they would have had if $I$ and $J$ had been precisely conserved. We conclude that the upper bound (\ref{A_CGL_subsonic}) survives, with $K_0$ replaced with $K_1$, the kinetic energy of the forced plasma flows.\footnote{It is not hard to see that the upper bound (\ref{CGL limit}) still holds in the somewhat more general case of $W$ changing by an order-unity (rather than small) amount if one can argue that the dynamics nevetheless preserves $I$ and $J$. The upper bound (\ref{CGL limit}) is only broken if the total energy increases by a large factor.} 

Obviously, the same line of argument can be used to extend to the forced case the arguments that we have proposed for kinetic MHD (section \ref{sec_A_KMHD}) and other kinetic models that conserve the magnetic moment of one of the bulk particle species (section \ref{sec_A_other}). 
                                                       
\subsection{Numerical tests}

To illustrate the above considerations of the dependence of dynamo action on the equation of state, we contrast a series of numerical simulations of the conventional MHD equations with isotropic pressure against the anisotropic-pressure CGL equations, both in the subsonically forced regime. Equations (\ref{conteq})-(\ref{Ampere}) are solved numerically in three spatial dimensions using a MUSCL-type scheme with a van Leer flux limiter \cite{Kurganov}. A divergence-free magnetic field is ensured by applying a flux-constrained approach (also known as a specific variation of staggered mesh or Yee grid method) \cite{Balsara}. Either the MHD energy equation (\ref{ideal MHD}) with isotropic pressure or the CGL equations (\ref{1})-(\ref{2}) are used. No resistive or viscous terms are included explicitly, so dissipation is due only to a small amount of numerical diffusion introduced by the numerical scheme. Periodic boundary conditions are applied in all three spatial directions. The simulation grid is uniform, with resolution of $128^3$ points (which is enough for the purposes of capturing dynamo action \cite{Meneguzzi,Haugen,Schekochihin-2004}). 

A stochastic flow in the simulation box is established by including a forcing term in the momentum equation \cite{Alvelius} that is uniform in wave-vector space and concentrated to the two smallest wave numbers. The forcing is random and white in time. Solenoidal ($\nabla \cdot {\bf F} = 0$) forcing is used because this provides the most effcient amplification of the initial magnetic field by the turbulent small-scale dynamo \cite{Federrath}. Note that, although the forcing is solenoidal, the equations that we solve allow for compressible dynamics and small ($\nabla \cdot {\bf V} \sim {\rm Ma}^2$) compressive fluctuations are present in the system. The injected power is constant in time and relatively small to ensure that the generated velocity fluctuations are subsonic. A small uniform mean seed magnetic field is set up in the initial condition,\footnote{In a periodic box, such a field cannot, of course, decay, and a certain amount of its amplification will be due just to the turbulent tangling of field lines, rather than to {\em bona fide} dynamo action \cite{Schekochihin-2007}. However, if the system does support dynamo action, the latter will quickly take over and bring the field energy to within a finite factor of the kinetic energy of the motions. Starting with small random magnetic perturbations whose mean is zero did not change any of our results.} and different strengths are used in the simulations described below, whereby the dependence on initial $\beta_0$ is investigated. All fields (velocity, density, magnetic field, perpendicular and parallel pressures) are initially constant across the simulation box, so their fluctuations are driven solely by forcing.

Since we are interested in the consequences of accurate conservation of $S$ in the case of isotropic pressure and $I$ and $J$ in the CGL case, we use either the isotropic-pressure relation (\ref{ideal MHD}) or the CGL equations (\ref{1})-(\ref{2}) instead of the total energy conservation equation in the explicit form.\footnote{Since our initial state is homogeneous, equations (\ref{ideal MHD}), (\ref{1}) or (\ref{2}) can be solved simply by enforcing pointwise conservation of $p/\rho^\gamma$, $p_\perp/\rho B$ or $p_\parallel B^2/\rho^3$, respectively.}. This implies that the part of the kinetic energy injected into the system that is dissipated numerically rather than being transformed into magnetic energy or into thermal energy via adiabatic mechanisms (compressional heating or, in CGL, pressure-anisotropy heating, i.e., parallel viscosity), is lost and thus constitutes an implicit energy sink in the total energy budget. As we argued in section \ref{sec_forced_MHD}, this is a small effect over a finite number of dynamical times. Various other schemes for solving our pressure equation(s) (enforcing total energy conservation or depositing the balance of numerically dissipated energy into $p_\parallel$ and/or $p_\perp$ according to physical assumptions about the nature of sub-grid heating) do not result in any change of the results on dynamo action or lack thereof reported below. 

\begin{figure}[t]
\begin{center}
\includegraphics[width=7cm]{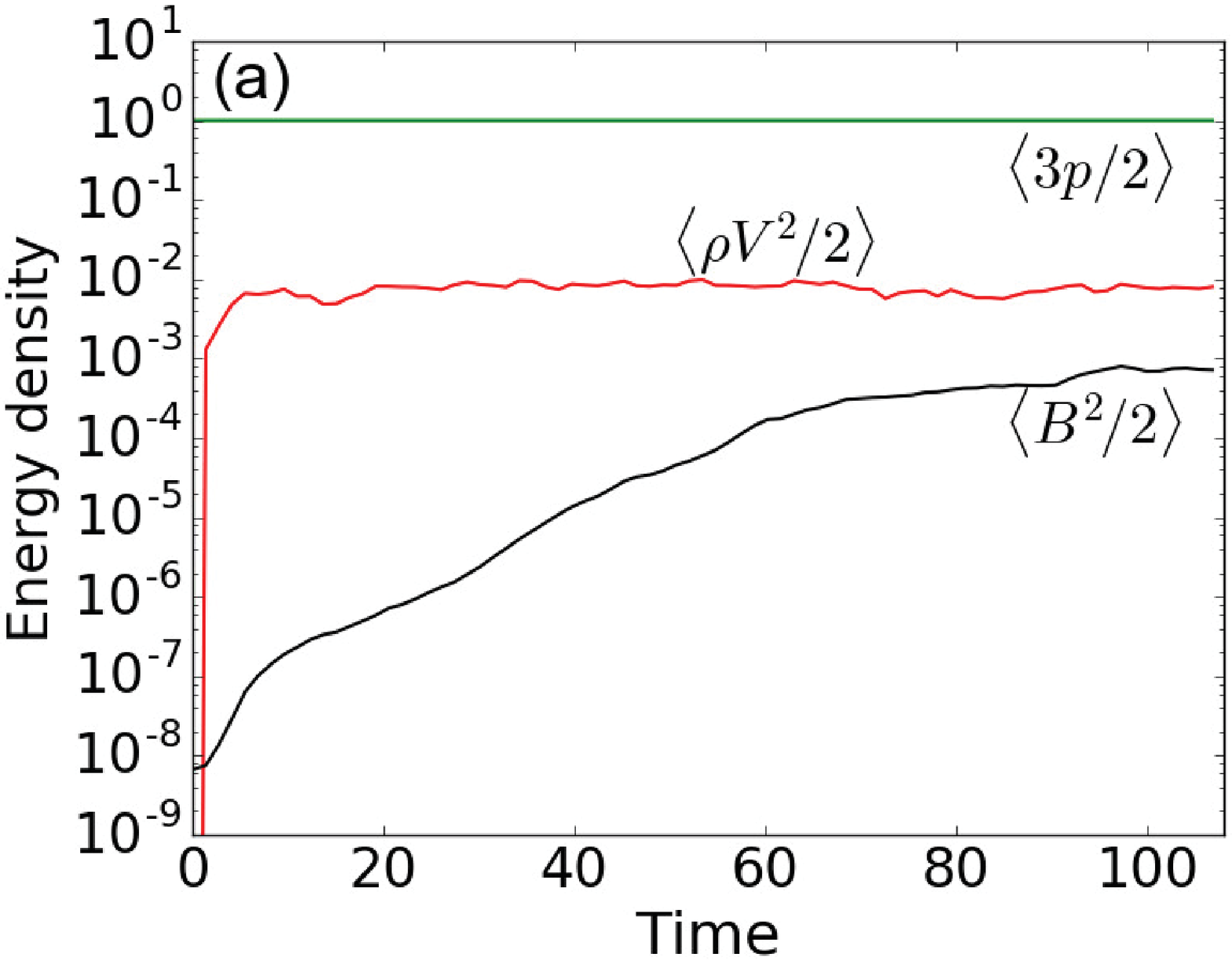}
\includegraphics[width=7cm]{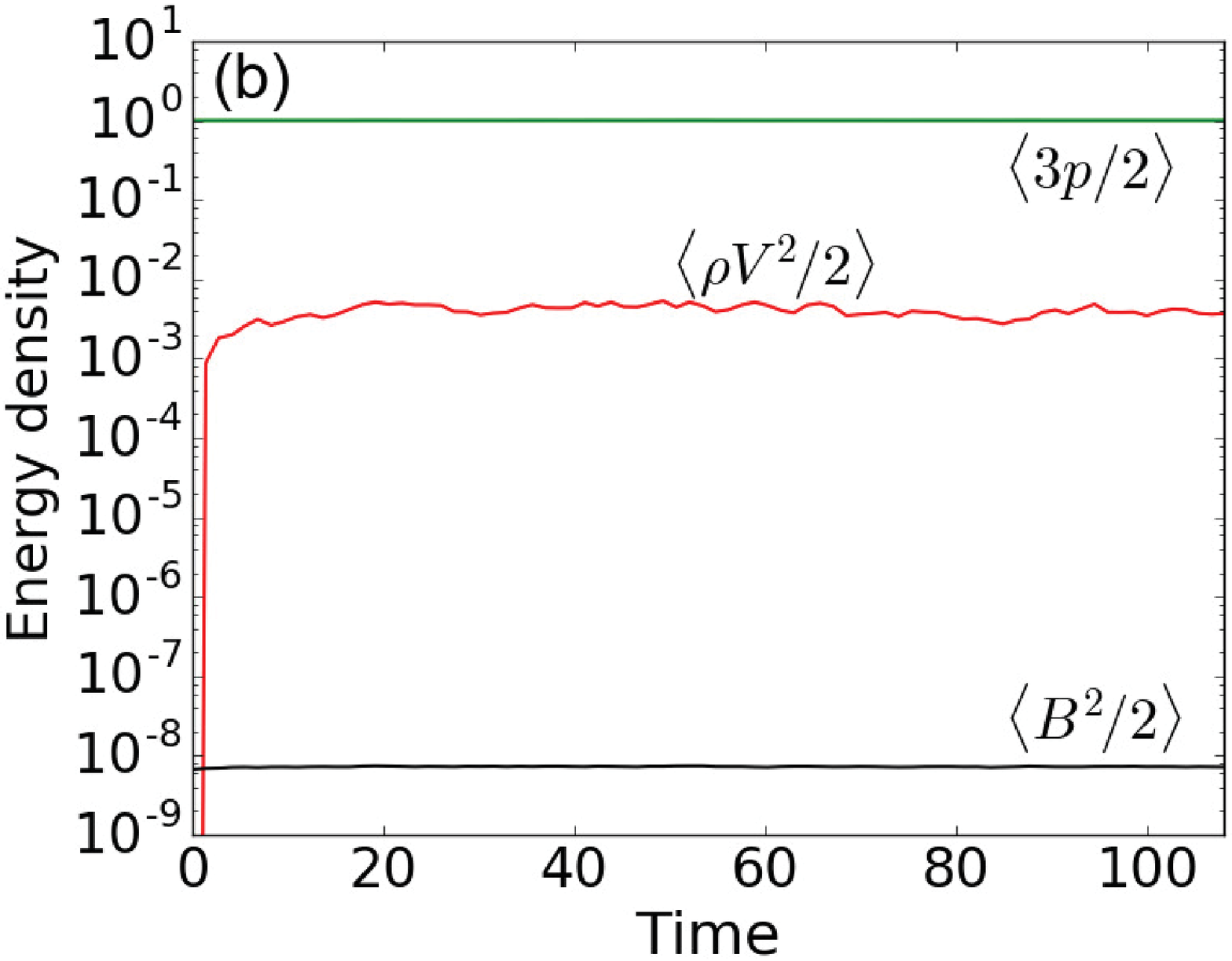}
\caption{Typical time evolution of the total thermal (green), kinetic (red) and magnetic (black) energies in (a) forced MHD equations with isotropic pressure and (b) forced CGL equations. The energies are in the units of the initial thermal energy $3p_0/2$. The time is in the units of the sound-crossing time $L/\sqrt{\gamma p_0/\rho_0}$, where $L=1$ is the size of the box. The largest-eddy turnover time, referred to as ``dynamical time'' in sections \ref{sec_forced_MHD} and \ref{sec_forced_CGL}, is approximately $\tau_{\rm dyn}\sim 10$ units. Note that the MHD dynamo operates on the time scale of order $\tau_{\rm dyn}$.}
\label{f1}
\end{center}
\end{figure}

\begin{figure}
\begin{center}
\includegraphics[width=7cm]{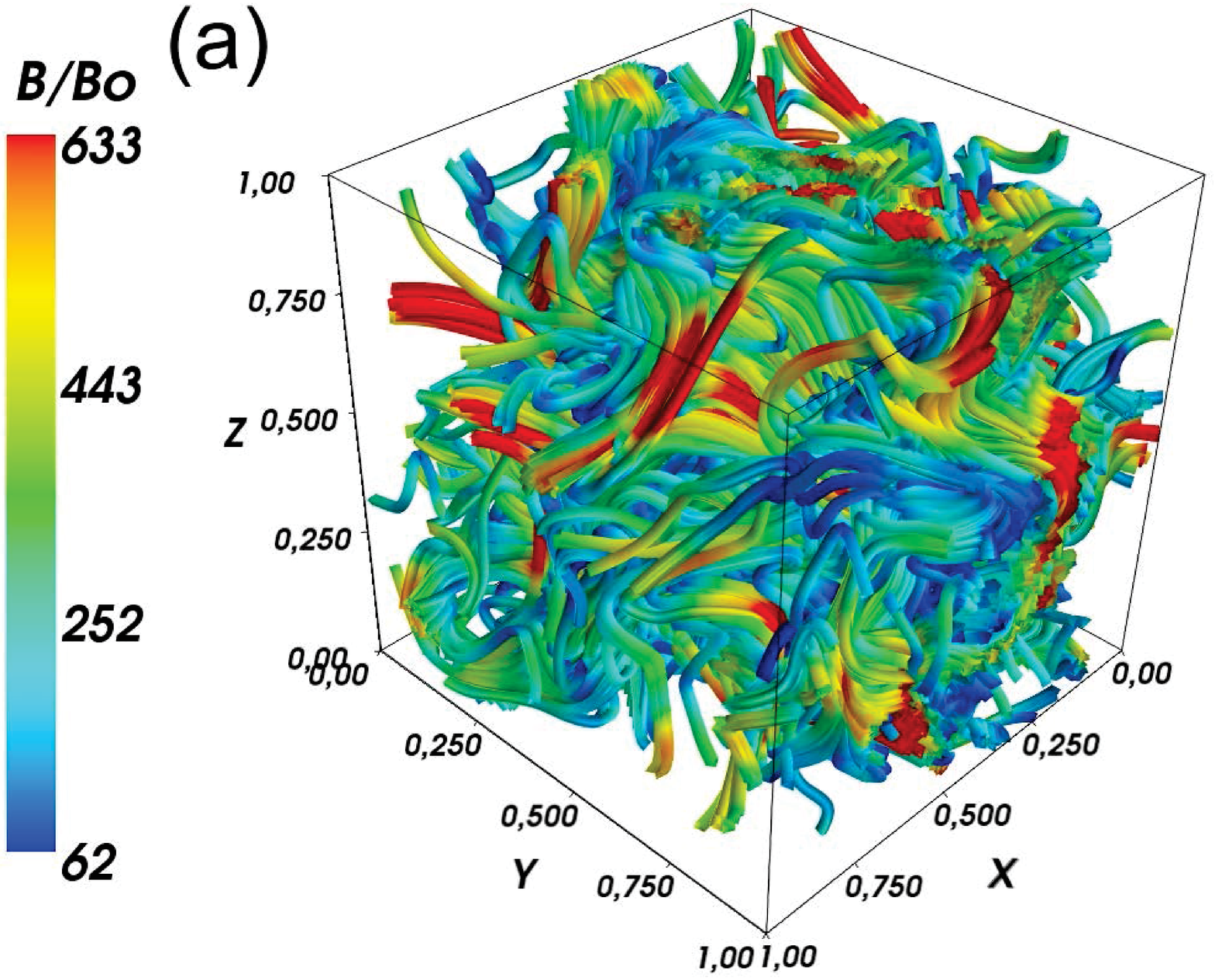}
\includegraphics[width=7cm]{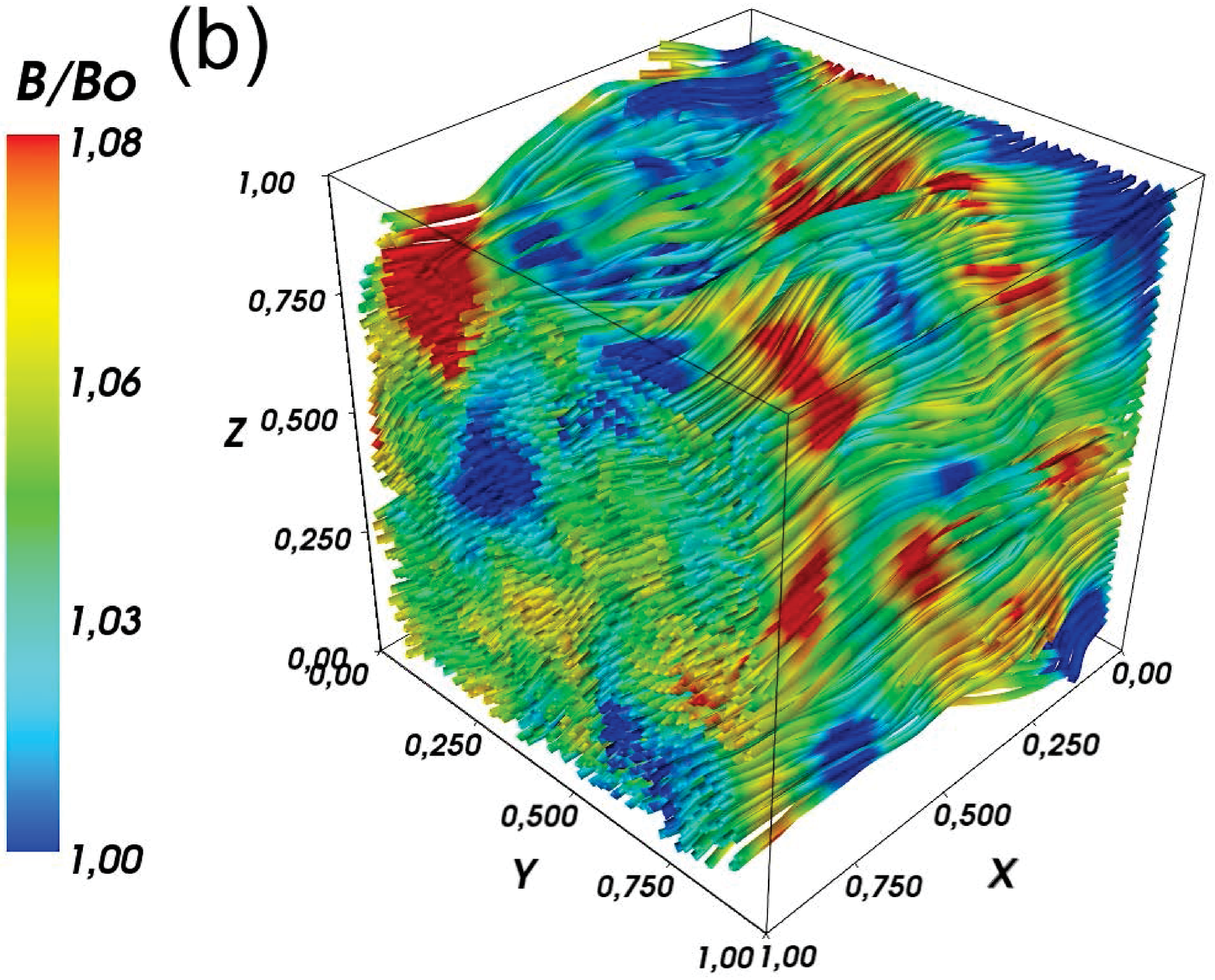}
\caption{Snapshots from the two simulations shown in Figure \ref{f1}, taken at $t=107$ (in the saturated state): (a) MHD, (b) CGL. 3D rendering of field lines is used to exhibit their spatial structure, while the colour shows the local value of the magnetic-field strength $B$ (normalised to the initial value $B_0$). The initial mean seed field was in the $x$ direction. The field lines that are displayed start at (a) $x = 0.5$, (b) $x = 0$.}
\label{f0}
\end{center}
\end{figure}

Figure \ref{f1} shows the evolution of the kinetic $\lang \rho V^2/2 \rang$, magnetic $\lang B^2/2 \rang$ and thermal $\lang 3 p/2 \rang$ energies in a simulation starting from $\beta_0 = 10^8$. In the MHD model, the magnetic energy grows exponentially until a saturated state is achieved, as it always does in dynamo simulations \cite{Meneguzzi,Haugen,Schekochihin-2004}. In contrast, in the CGL model, the dynamo action is suppressed, and no significant increase of the magnetic energy density is obtained, confirming the result of \cite{Lima}. 
A 3D rendering of the magnetic field in the saturated state of these two simulations is given in Figure \ref{f0}. In the isotropic-pressure MHD plasma, plasma motions were able to amplify the magnetic field via the standard stretch-and-fold mechanism, giving rise to a characteristic pattern of intertwined and folded magnetic field lines and to intermittent high- and low-field regions. The contrast between the highest- and the lowest-$B$ regions is of order of 10, and even the lowest-magnitude regions have been amplified by a factor of $\sim60$ from the initial field. In the CGL case, the same initial condition and forcing resulted in very weak bending of the lines and very small amplification of the magnetic field strength (a few per cent), the spatial alignment of the lines displays no signature of stretching or folding, just gentle wave-like perturbations (these are CGL slow waves, which propagate at the sound speed even at high $\beta$ \cite{Shrauner}). 

\begin{figure}[t]
\begin{center}
\includegraphics[width=7cm]{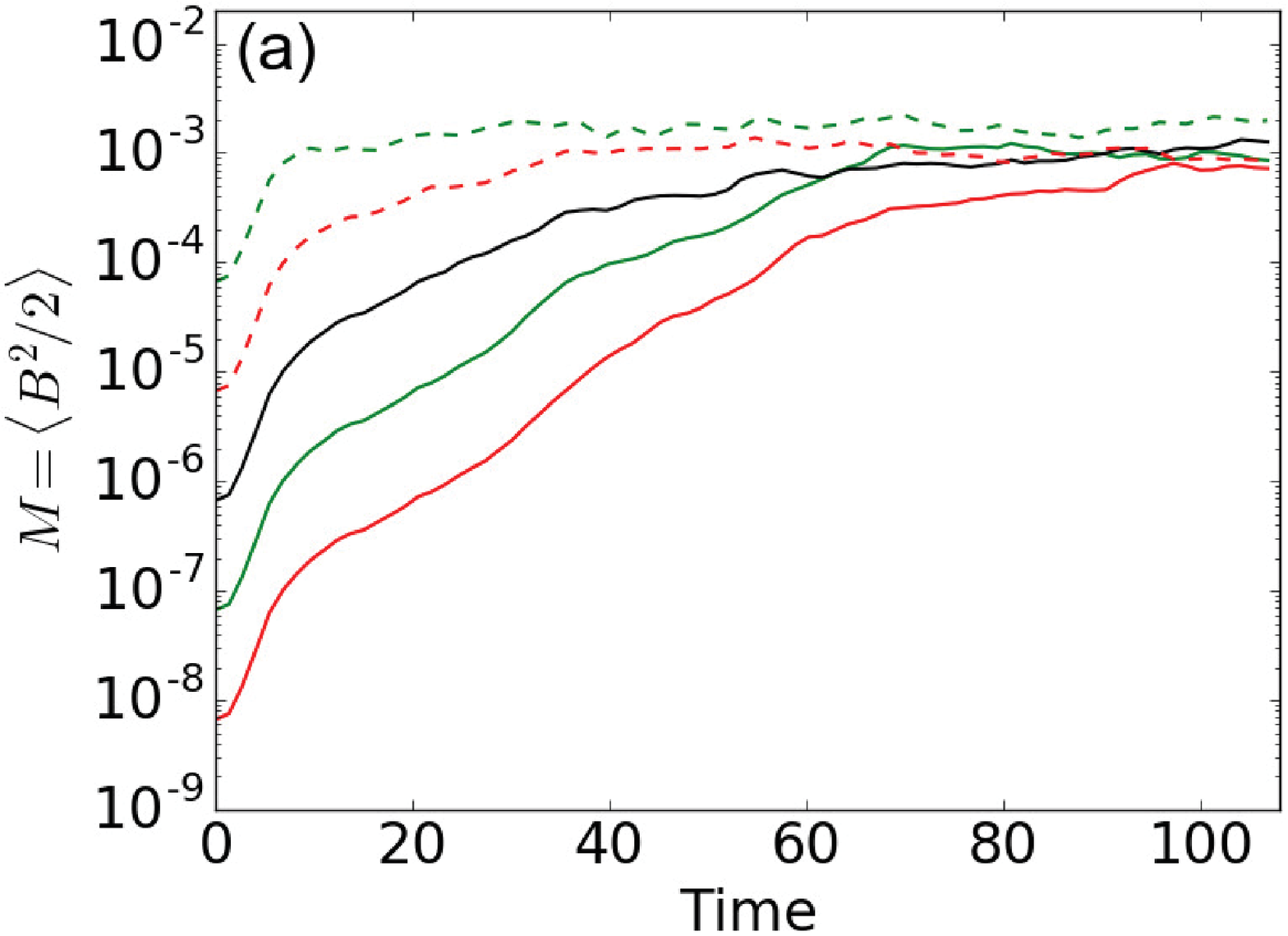}
\includegraphics[width=7cm]{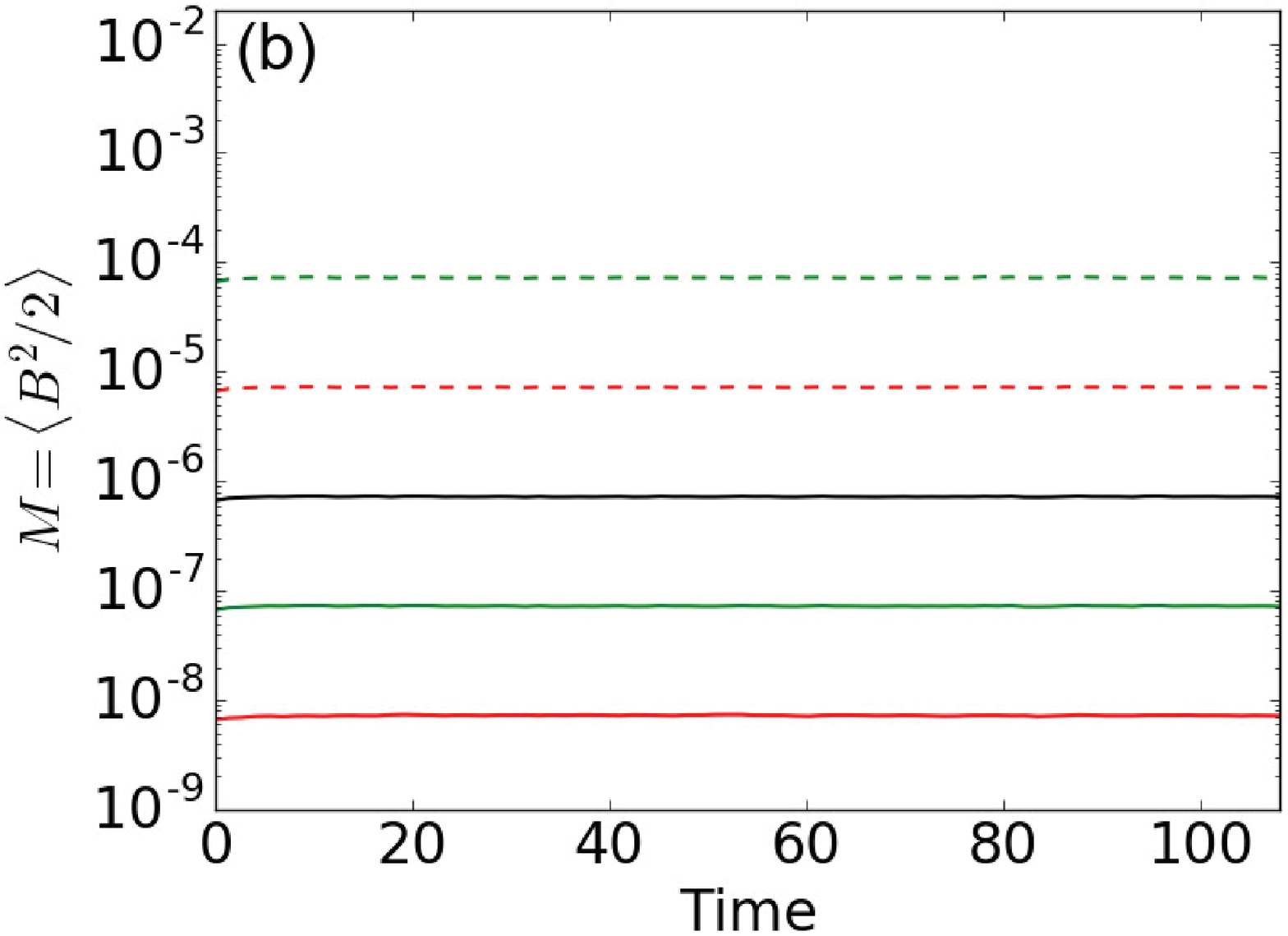}
\caption{Time evolution of the magnetic energy  
for (a) the MHD model with isotropic pressure and (b) the CGL model, 
for five runs starting with different initial magnetic fields. The units are 
the same as in Figure \ref{f1}.}
\label{f2}
\end{center}
\end{figure}

\begin{figure}[h!]
\begin{center}
\includegraphics[width=7cm]{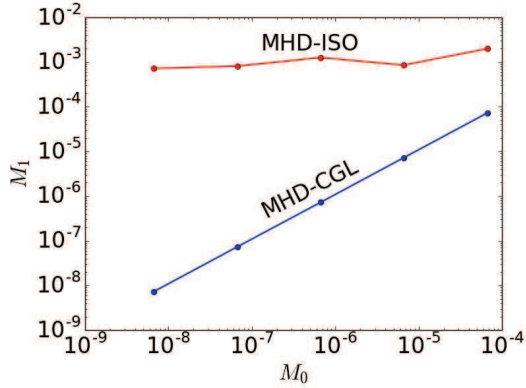}
\caption{The saturated magnetic energy $M_1$ vs.\ the initial one $M_0$ 
for the MHD (red) and CGL (blue) models.}
\label{f3}
\end{center}
\end{figure}

Figure \ref{f2} show the evolution of the magnetic energy for the two models in simulations starting from the magnetic field strength $B_0$ corresponding to the range $10^4 < \beta_0 < 10^8$. The amplitude of the forcing and the thermal pressure are the same for all these simulations, implying approximately the same Mach number of the velocity fluctuations in all of them. We see that the saturated magnetic energy is essentially independent of $\beta_0$ in the MHD model and entirely determined by it in the CGL model. This point is reinforced by Figure \ref{f3}, which shows the final saturated magnetic energy $M_1$ vs.\ the initial one $M_0$. Thus, the initial magnetic energy cannot be increased in a double-adibatic, pressure-anisotropic plasma even in the presence of external forcing and the upper bound (\ref{CGL limit}) is tight.

\section{Conclusions}

As we have seen, any mathematical plasma model that conserves the magnetic moment of at least one particle species is subject to an anti-dynamo theorem in the sense that there is a strict limitation on the growth of the total magnetic energy, as indeed anticipated by \cite{Kulsrud-1997}. If the initial magnetic field is small enough that $\beta_0 \gg 1$ (or $\beta_0 \gg {\rm Ma}^{-2}$ for plasma flows in which the Mach number is small), the magnetic energy cannot grow by more than a factor of order unity as long as the external forcing does not significantly increase the total energy of the system.

These results have immediate implications for dynamo simulations of the collisionless CGL or kinetic-MHD equations. If such a simulation should exhibit significant growth of the magnetic energy, it must be due to the accumulation of numerical errors that effectively break the adiabatic invariance of the plasma. Our own simulations of the CGL equations, as well as those of \cite{Lima}, find that dynamo action is, indeed, practically absent at high beta. Moreover, the modest growth of the magnetic energy that does occur is observed to stop at magnetic energy inversely proportional to $\beta_0$, as expected from the bound (\ref{CGL limit}), suggesting that this scaling is the correct one and that the bound is tight within a factor of order unity. In simulations that instead use the conventional MHD equation of state (\ref{ideal MHD}), the magnetic energy grows to much larger values, independent of the inital beta. 

These results do not, however, imply the absence of dynamo action in a real plasma, even if collisions are rare and the gyroradii of all particle species are small. There are at least two reasons why such a conclusion cannot be drawn. First, whether or not external forcing is present, the flow of the plasma will usually be turbulent. The ensuing free-energy cascade will tend to create small-scale structures both in real space and in velocity space, with fluctuations arising on Larmor scales and collisions eventually becoming important \cite{Schekochihin-2008}. The precise adiabatic invariance of the magnetic moment will thus be broken, which may unchain the dynamo. Secondly, the growth of pressure anisotropies caused by magnetic-field changes can lead to the excitation of kinetic mirror and firehose instabilities that have a similar effect \cite{Kunz,Melville}. Without a detailed understanding of these processes, it is not possible to rule out dynamo action. Indeed, it would be unwise to do so, given that the Universe {\em is} magnetised, the observed magnetic fields tend to be of dynamical strength (and so likely result from plasma motions), and first numerical evidence of collisionless plasma dynamo action has appeared \cite{Rincon}. What we can conclude with certainty is that plasma dynamo action must be a multiscale process, with breaking of adiabatic invariance at microscales playing an existentially important role. 

\subsection*{Acknowledgment} The work of AAS was supported in part by grants from the UK STFC and EPSRC. MS was supported
by funding from the European Union's Horizon 2020 research and
innovation programme under the Marie Sklodowska-Curie grant agreement No
657251 (ASTROMULTISCALE). The discussion presented in the paper reflects
only the authors' view and the European Commission is not responsible for
any use that may be made of the information it contains. Numerical simulations
presented in this paper
were performed on supercomputers of the Academic Computer Centre in Gdansk
(CI TASK).

\section*{Appendix A: Invariants of the CGL equations}

The double-adiabatic equations of state can be expressed as
	$$ \frac{d}{dt} \left[ \alpha \ln  \left( \frac{p_\perp}{\rho B} \right) + (1-\alpha) \ln \left( \frac{p_\|^{1/3} B^{2/3}}{\rho} \right) \right] = 0, $$
where $\alpha$ is arbitrary. Since $d \ln \rho / dt = - \nabla \cdot {\bf V}$ from the continuity equation, we can thus write
	$$ \frac{d}{dt} \ln \left( p_\perp^\alpha p_\|^{(1-\alpha)/3} B^{(2-5 \alpha)/3} \right) + \nabla \cdot {\bf V} = 0, $$
and conclude that there is an infinity of conserved quantities that can be constructed from the three fields $p_\perp$, $p_\|$ and $B$, namely, 
	$$ \frac{d}{dt} \lang p_\perp^\alpha p_\|^{(1-\alpha)/3} B^{(2-5 \alpha)/3} \rang = 0. $$
Three of these invariants depend on only two of the fields: $I = \lang p_\perp/B \rang$ and $J= \lang p_\|^{1/3} B^{2/3} \rang$ correspond to $\alpha = 1$ and $\alpha = 0$, respectively, and $\alpha = 2/5$ yields an invariant that is independent of $B$, 
	$$ \frac{d}{dt} \lang p_\perp^{2/5} p_\|^{1/5}\rang = 0. $$

\end{document}